\documentclass[onecolumn]{template}    

\usepackage{graphicx}         
\usepackage{amsmath}
\usepackage{amsfonts}
\usepackage{enumerate}   
\usepackage{bm}   
\usepackage{amssymb}
\usepackage{chngcntr}
\usepackage{ dsfont }
\usepackage{apptools}
\usepackage[dvipsnames]{xcolor}
\usepackage{color}
\usepackage{hyperref}

\newtheorem{example}{Example}

\newcommand{\mf}{\mathbf}

\newcommand{\sgn}[2]{\left\lceil#1\right\rfloor^{#2}}
\newcommand{\diag}[1]{\mbox{diag}(#1)}

\def\RevEight#1{\textcolor{black}{#1}}
\def\RevTwo#1{\textcolor{black}{#1}}
\def\RevFour#1{\textcolor{black}{#1}}
\def\RevAll#1{\textcolor{black}{#1}}

\begin{document}

\begin{frontmatter}

\title{REDCHO: \\Robust Exact Dynamic Consensus of High Order} 

\thanks[footnoteinfo]{\textcolor{red}{This is the preprint version of the accepted Manuscript: Rodrigo Aldana-López, Rosario Aragüés, Carlos Sagüés,
“REDCHO: Robust Exact Dynamic Consensus of High Order”, Automatica, Volume 141, 2022, ISSN 0005-1098. DOI: 10.1016/j.automatica.2022.110320.
Please cite the publisher's version. For the publisher's version and full citation details see:
\url{https://doi.org/10.1016/j.automatica.2022.110320}\\
Simulation files for the algorithms presented in this work can be found on \url{https://github.com/RodrigoAldana/EDC}
}\\This work was supported by projects COMMANDIA SOE2/P1/F0638
(Interreg Sudoe Programme, ERDF), PGC2018-098719-B-I00 (MCIU/ AEI/
FEDER, UE) and DGA / FSE T45\_20R(Gobierno de Aragon).  The authors would like to acknowledge the sponsorship of Universidad de Zaragoza, Banco Santander and CONACYT, México.}

\author[First]{Rodrigo Aldana-López*} 
\author[First]{Rosario Aragüés} 
\author[First]{Carlos Sagüés} 

\address[First]{Departamento de Informatica e Ingenieria de Sistemas (DIIS) and Instituto de Investigacion en Ingenieria de Aragon (I3A), 
\\
Universidad de Zaragoza, Zaragoza 50018, Spain.\\
(e-mail: rodrigo.aldana.lopez@gmail.com, raragues@unizar.es, csagues@unizar.es)}

\begin{abstract}               
This article addresses the problem of average consensus in a multi-agent system when the desired consensus quantity is a time varying signal. Recently, the EDCHO protocol leveraged high order sliding modes to achieve exact consensus under a constrained set of initial conditions, limiting its applicability to static networks. In this work, we propose REDCHO, an extension of the previous protocol which is robust to mismatch in the initial conditions, making it suitable to use cases in which connection and disconnection of agents is possible. The convergence properties of the protocol are formally explored. \RevAll{Finally, the effectiveness and advantages of our proposal are shown with concrete simulation examples showing the benefits of REDCHO against other methods in the literature.}
\end{abstract}


\end{frontmatter}

\section{Introduction}

In recent decades, the problem of dynamic consensus have received a lot of attention. The goal of this problem is to make a team of agents agree in a time-varying quantity through a distributed algorithm, by sharing information with their neighbors in a communication network. Some examples of applications of dynamic consensus include distributed estimation in sensor networks \cite{freeman2010}, distributed convex optimization \cite{Kia2015b}, distributed coordination in electrical grids \cite{Cherukuri2016} and distributed formation control \cite{porfiri2007}.

\RevAll{{In distributed average tracking (DAT) applications, dynamic consensus algorithms are used as virtual observers for the average of some time-varying reference signals, which can be tracked by a controller for a local physical system at each agent \cite{zhao2019,sen2020}. Depending on the order of the system it may be desirable for the dynamic consensus observer to obtain derivatives of the average signal \cite{sen2020}}. In this context, the works \cite{freeman2006,Solmaz2017,Olfati2007} propose linear dynamic consensus algorithms for scalar systems. However, these algorithms have the disadvantage of having a non-zero terminal error bound for some classes of reference signals. This issue has been tackled for scalar systems in  \cite{nostrati2012,freeman2019} and second order systems in \cite{Chen2012,ghan2019,zhao2019} by means of First Order Sliding Modes (FOSM), allowing exact convergence for more general classes of reference signals. However, these approaches suffer from the so-called chattering effect due to the discontinuous character of the FOSM \cite[Chapter 3]{fridman2002}. This makes the system sensitive to delays and noise. For the high-order case, some algorithms are able to obtain the average signal and its derivatives with exact convergence for vanishing reference differences in \cite{sen2020} and reference differences with a bounded high-order derivative in \cite{zhao2017}. Nonetheless, these approaches impose a higher communication burden since agents share all high-order errors instead of a single scalar. On the other hand, The Exact Dynamic Consensus of High Order (EDCHO) algorithm from \cite{aldana2021} manages to obtain the average signal and its high-order derivatives when the agents share a single scalar and the reference signals have differences with some bounded high-order derivative. Moreover, it employs employs High Order Sliding Modes (HOSM) which mitigates the chattering effect of FOSM algorithms. }

The main limitation of EDCHO is that the initial values of the agent states must be constrained within a specific surface. A similar condition is also required in \cite{chen2015,zhao2019,zhao2017,sen2020}.\RevTwo{ This issue dramatically limits its applicability since this initial requirement can be violated when agents connect or disconnect from the network, preventing the protocol from converging in an uncertain or time-varying environments as in the real network scenarios described in \cite{Zhang2013}.} The mismatch in the initial conditions has been tackled for linear algorithms in \cite[Page 57]{Solmaz2017} and for the robust FOSM algorithm in \cite{freeman2019}. However, the extension of these techniques to EDCHO is non-trivial due to its high-order non-linear system character.

Motivated by the previous discussion, we propose Robust EDCHO (REDCHO), which manages to achieve dynamic consensus with exact convergence towards the average signal and its high-order derivatives, regardless of isolated events of spontaneous  changes in the network. \RevTwo{This feature improves the applicability of REDCHO in uncertain or time-varying networks with respect to \cite{aldana2021}. In addition, the use of HOSM in REDCHO mitigates the effect of chattering when compared with \cite{freeman2019}. \RevFour{Moreover, REDCHO allows the agents to share only a scalar, reducing the communication burden compared to \cite{sen2020,zhao2017}.} }
\subsection{Notation}

Let $\mathds{1}_n = [1,\dots,1]^T\in\mathbb{R}^n$ and $\text{\normalfont I}_n\in\mathbb{R}^{n\times n}$ the identity matrix.  Let $\text{diag}(v), v\in\mathbb{R}^n$ represents a diagonal matrix whose diagonal is composed by $v$. Let $\otimes$ denote the Kronecker product Let $\text{sign}(x) = 1$ if $x> 0, \text{sign}(x)=-1$ if $x<0$ and $\text{sign}(0)=0$. Moreover, if $x\in\mathbb{R}$, let $\lceil x\rfloor^\alpha:= |x|^\alpha\text{sign}(x)$ for $\alpha>0$ and $\lceil x\rfloor^0:={\text{sign}}(x)$. When $x=[x_1,\dots,x_n]^T\in\mathbb{R}^n$, then $\sgn{x}{\alpha}:=\left[\sgn{x_1}{\alpha},\dots,\sgn{x_n}{\alpha}\right]^T$ for $\alpha\geq 0$. If $V:\mathbb{R}^n\to\mathbb{R}$ is continuously differentiable, denote $\frac{\partial V}{\partial x} = \left[\frac{\partial V}{\partial x_1},\dots,\frac{\partial V}{\partial x_n}\right]$. Furthermore, let $V:\mathbb{R}^n\to\mathbb{R}$ be continuous and $F:\mathbb{R}^n\to\mathbb{R}^n$, then $L_FV:\mathbb{R}^n\to\mathbb{R}$ denotes the Lie derivative. \RevTwo{Finally, let $x^{(\mu)}(t)$ denote the $\mu$-th derivative of a signal $x(t)$.}

\section{Problem statement}

Consider a multi-agent system distributed in a network $\mathcal{G}$. In the following $\mathcal{G}$ is an undirected connected graph of $n$ nodes characterized by its incidence matrix $D$ or its adjacency matrix $A$ \cite[Chapter 8]{Godsil}. We consider that each agent $i$ has access to a local time varying signal $u_i(t)\in\mathbb{R}$. Additionally, each agent is capable of communicating with their neighbors \RevFour{according to a communication topology defined by $\mathcal{G}$}.  Moreover, each agent $i$ \RevTwo{runs a local observer with} output $y_i(t)=[y_{i,0}(t),\dots,y_{i,m}(t)]^T\in\mathbb{R}^{m+1}$ \RevFour{where $m$ is a desired system order}. \RevFour{In order to reduce communication burden, we require all agents to share only $y_{i,0}$ instead of the whole $y_i(t)$}. The goal of the system is to achieve the following property.

\RevFour{
\begin{defn}\upshape{\bf Robust Exact Dynamic Consensus.} Let the average signal $
\bar{u}(t) := \frac{1}{n}\sum_{i=1}^nu_i(t)$. Then, the multi-agent system is said to achieve robust EDC, if the individual output signals for each agent comply
\begin{equation}
\label{eq:prob}
    \begin{aligned}
    &\lim_{t\to\infty}\left|y_{i,\mu}(t) - \bar{u}^{(\mu)}(t)\right| = 0,
    \end{aligned}
\end{equation}
for all $\mu\in\{0,\dots,m\}, i\in\{1,\dots,n\}$ regardless of  isolated events of spontaneous connection or disconnection of agents.
\end{defn}}
\vspace{-0.8cm}
\RevAll{\begin{rem}
Assume that all agents are provisioned with an observer complying the properties described before. Moreover, assume all agents have a local physical system with dynamics of relative degree $m$. Then, the local observer's output $y_i(t)$ can be used locally at each agent to solve a DAT problem. In this case, a local feedback control can be designed using the ideas from \cite{sen2020,ghan2019,zhao2019}.
\end{rem}
}
\section{REDCHO}
We propose a new algorithm, REDCHO, which manages to achieve robust EDC under mild assumptions on the reference signals. To present the algorithm, first let the auxiliary matrices
$$
\begin{aligned}
&\Gamma = 
\begin{bmatrix}
-\gamma_0 & 1 & 0 &\cdots&0 \\
0 &-\gamma_1 & 1 & \ddots&\vdots \\\
\vdots&\ddots&\ddots&\ddots&0\\
\vdots& &\ddots&-\gamma_{m-1}&1 \\
0&\cdots&\cdots&0&-\gamma_{m}
\end{bmatrix},
G = \begin{bmatrix}
C\\
C\Gamma \\
\vdots \\
C\Gamma^{m}
\end{bmatrix}
\end{aligned}
$$
for design parameters $\gamma_0,\dots,\gamma_m>0$ and \RevFour{$C = [1,0,\dots,0]\in\mathbb{R}^{1\times(m+1)}$}. The structure of the REDCHO algorithm is proposed as
\RevFour{
\begin{equation}
\label{eq:redcho}
\begin{array}{ll}
    \dot{x}_{i,\mu}(t) &= k_\mu\theta^{\mu+1} \sum_{j=1}^na_{ij}\lceil y_{i,0}(t) - y_{j,0}(t) \rfloor^{\frac{m-\mu}{m+1}}\\&+\ x_{i,\mu+1}(t) - \gamma_\mu x_{i,\mu}(t),\ \ \ 0\leq \mu < m\\
    \dot{x}_{i,m}(t) &= k_m\theta^{m+1} \sum_{j=1}^na_{ij}\sgn{y_{i,0}(t) - y_{j,0}(t) }{0} \\&- \gamma_mx_{i,m}(t)\\
    y_{i,\mu}(t) &= u_i^{(\mu)}(t) - \sum_{\nu=0}^{m} G_{\mu+1,\nu+1} x_{i,\nu}(t).
\end{array}
\end{equation}
}where $G_{\mu+1,\nu+1}$ with $\mu,\nu\in\{0,\dots,m\}$ are the components of the matrix $G$ \RevFour{and $\theta\geq 1$ is a design parameter}. Furthermore, we consider the following assumption:
\begin{assum}
\label{as:u_signals}
Let 
$w_i(t)=\left(\bar{u}^{(m+1)}(t) - {u}_i^{(m+1)}(t)\right) + \sum_{\mu=0}^m l_\mu \left(\bar{u}^{(\mu)}(t)-{u}_i^{(\mu)}(t)\right)$
where $l_0,\dots,l_m$ are the coefficients of the polynomial $(s+\gamma_0)\cdots(s+\gamma_m) = s^{(m+1)} + \sum_{\mu=0}^{m}l_\mu s^{\mu}$.
Thus, $|w_i(t)|\leq L, \forall t\geq t_0$ \RevFour{for fixed $\gamma_0,\dots,\gamma_m$ and known $L>0$}.
\end{assum}

{It is easy to show that the EDCHO algorithm from \cite{aldana2021} is a particular limiting case of REDCHO and can be recovered by choosing $\gamma_0=\cdots=\gamma_m=0$ \RevFour{and $\theta=1$}. However, EDCHO assumes that $\sum_{i=1}^nx_{i,\mu}(t_0)= 0$ is satisfied. This condition breaks easily, specially when agents connect or disconnect from the network. }The main result of this work, which is formally stated and shown in Section \ref{sec:main_result}, is that using the gains $k_0,\dots,k_m$ designed as in \cite[Theorem 7]{aldana2021} and under Assumption \ref{as:u_signals}, then REDCHO algorithm works even when $\sum_{i=1}^nx_{i,\mu}(t_0)\neq 0$ and thus achieves robust EDC. In order to formally show these facts, we provide some auxiliary results.

\section{Towards convergence of REDCHO}
The REDCHO algorithm \eqref{eq:redcho} can be written in partially vectorized form as:
\RevFour{
\begin{equation}
\begin{aligned}
\label{eq:main_algo_vec2}
    \dot{X}_{\mu}(t) &= X_{\mu+1}(t) + k_\mu\theta^{\mu+1} D\sgn{D^TY_{0}(t)}{\frac{m-\mu}{m+1}} - \gamma_\mu X_\mu(t)\\ &\text{for } 0\leq\mu < m, \\
    \dot{X}_{m}(t) &= k_m\theta^{m+1}D\sgn{D^TY_0(t)}{0} - \gamma_m X_m(t)\\
    Y_\mu(t) &= U^{(\mu)}(t)-\sum_{\nu=0}^{m} G_{\mu+1,\nu+1}X_{\nu}(t), \forall \mu\in\{0,\dots,m\}
\end{aligned}
\end{equation}
}
where we define $X_\mu(t):=[x_{1,\mu}(t),\dots,x_{n,\mu}(t)]^T, Y_\mu(t):=[y_{1,\mu}(t),\dots,y_{n,\mu}(t)]^T$, $U(t):=[u_1(t),\dots,u_n(t)]^T$ and \RevTwo{$D$ is the incidence matrix of $\mathcal{G}$}. Moreover, let $\mf{X}(t):=[X_0(t)^T,\dots,X_m(t)^T]^T, \mf{Y}(t):=[Y_0(t)^T,\dots,Y_m(t)^T]^T$, $\mf{U}(t):=\left[\left(U^{(0)}(t)\right)^T,\dots,\left(U^{(m)}(t)\right)^T\right]^T$ and
\RevFour{
$$
\mf{F}(Y_0(t);\theta) = \begin{bmatrix}
k_0\theta D\sgn{D^TY_{0}(t)}{\frac{m}{m+1}}\\
\vdots\\
k_m\theta^{m+1}D\sgn{D^TY_{0}(t)}{0}
\end{bmatrix}
$$
}
Using this notation we obtain the fully vectorized form of the algorithm:
\begin{equation}
\begin{aligned}
\label{eq:main_algo_vec3}
    \dot{\mf{X}}(t) &= (\Gamma\otimes \text{\normalfont I}_n)\mf{X}(t) + \mf{F}(Y_0(t);\theta)\\
    \mf{Y}(t) &= \mf{U}(t)- (G\otimes \text{\normalfont I}_n)\mf{X}(t)
\end{aligned}
\end{equation}
Both partial and fully vectorized versions of the algorithm \RevFour{will be used throughout this work}. Moreover, note that $G$ is the observability matrix of the pair $(\Gamma,C)$ which is invertible. Then, the dynamics of $\mf{Y}(t)$ result in
\begin{equation}
\label{eq:dyn_y}
    \begin{aligned}
    \dot{\mf{Y}} &= \dot{\mf{U}} - (G\otimes \text{\normalfont I}_n)(\Gamma\otimes \text{\normalfont I}_n)\mf{X} - (G\otimes \text{\normalfont I}_n)\mf{F}(Y_0;\theta) \\
    &=\dot{\mf{U}} + (G\otimes \text{\normalfont I}_n)(\Gamma\otimes \text{\normalfont I}_n)(G^{-1}\otimes \text{\normalfont I}_n)(\mf{Y}-\mf{U})\\& + (G\otimes \text{\normalfont I}_n)\mf{F}(Y_0;\theta) \\
    &=\dot{\mf{U}} + (G\Gamma G^{-1}\otimes \text{\normalfont I}_n)(\mf{Y}-\mf{U}) - (G\otimes \text{\normalfont I}_n)\mf{F}(Y_0;\theta)
    \end{aligned}
\end{equation}
since $\mf{X}(t) = -(G^{-1}\otimes \text{\normalfont I}_n)(\mf{Y}(t)-\mf{U}(t))$, and $(G\otimes \text{\normalfont I}_n)(\Gamma\otimes \text{\normalfont I}_n)(G^{-1}\otimes \text{\normalfont I}_n) = (G\Gamma G^{-1}\otimes \text{\normalfont I}_n)$.

We will show that $\mf{Y}(t)$ converges towards the average consensus vector $\left[\bar{u}(t)\mathds{1}_n^T,\dots,\bar{u}^{(m)}(t)\mathds{1}_n^T\right]^T$ asymptotically achieving EDC. This analysis is performed by decomposing $\mf{Y}(t)=(\text{\normalfont I}_{m+1}\otimes\mathds{1}_n)\bar{y}(t)+\mf{\tilde{Y}}(t)\in\mathbb{R}^{(m+1)n}$ in the consensus component $\bar{y}(t) = (\text{\normalfont I}_{m+1}\otimes \mathds{1}_n^T/n)\mf{Y}(t)\in\mathbb{R}^{m+1}$ and in the consensus error $\tilde{\mf{Y}}(t) = (\text{\normalfont I}_{m+1}\otimes P)\mf{Y}(t)\in\mathbb{R}^{(m+1)n}$ with $P=(\text{\normalfont I}_n-(1/n)\mathds{1}_n\mathds{1}_n^T)$. Therefore, convergence of $\mf{Y}(t)$ can be established by means of showing that $\bar{y}(t)$ converges exponentially to $[\bar{u}(t),\dots,\bar{u}^{(m)}(t)]^T$ and $\tilde{\mf{Y}}(t)$ converges in finite time to the origin as we do in the following sections. First, we provide some results regarding structural properties of the matrices $\Gamma, G$ and their relation to the signals $u_i(t)$ and $w_i(t)$ from Assumption \ref{as:u_signals}. These notions will be useful in subsequent proofs.

\begin{lem}
\label{le:w_dynamics}
Let the change of variables $\mf{W}(t)=[W_0^T,\dots,W_m^T]^T = (G^{-1}\otimes \text{\normalfont I}_n)\mf{U}(t)$ with $W_\mu(t)\in\mathbb{R}^n, \forall \mu\in\{0,\dots,m\}$. Moreover, define 
\begin{equation}
\label{eq:Wmp1}
W_{m+1}(t) = U^{(m+1)}(t)+\sum_{\mu=0}^ml_\mu U^{(\mu)}(t)
\end{equation}
where $l_0,\dots,l_m$ are the coefficients of the polynomial $(s+\gamma_0)\cdots(s+\gamma_m) = s^{(m+1)} + \sum_{\mu=0}^{m}l_\mu s^{\mu}$. Then, we conclude that 
\begin{equation}
\label{eq:dynU}
\dot{\mf{U}} = (\tilde{\Gamma}\otimes \text{\normalfont I}_n)\mf{U} + (B\otimes \text{\normalfont I}_n)W_{m+1}
\end{equation}
and
\begin{equation}
\label{eq:dynW}
\dot{\mf{W}} = (\Gamma\otimes \text{\normalfont I}_n)\mf{W} + (B\otimes \text{\normalfont I}_n)W_{m+1}
\end{equation}
with $B=[0,\dots,0,1]^T\in\mathbb{R}^{(m+1)\times 1}$ and
\begin{equation}
\label{eq:companion}
\tilde{\Gamma} = \begin{bmatrix}
0 & 1 & 0 &\cdots& 0 \\
\vdots & 0 & \ddots & \ddots &\vdots \\
\vdots&\vdots&\ddots& \ddots &0 \\
0 & 0&\cdots& 0& 1 \\
-l_0 & -l_1 & \cdots & \cdots&-l_m
\end{bmatrix}.
\end{equation}
\end{lem}
\vspace{1em}
\begin{pf}

First, let $U_\mu(t):=U^{(\mu)}(t)$ and note that $\dot{U}_\mu(t) = U_{\mu+1}(t)$ for $0\leq \mu<m$ and $\dot{U}_m(t) = W_{m+1}(t) - \sum_{\mu=0}^ml_\mu U_\mu(t)$ from the definition in \eqref{eq:Wmp1}. Writing this in complete vector form leads to \eqref{eq:dynU} directly. Now, rewrite \eqref{eq:Wmp1} as
\begin{equation}
\begin{aligned}
\label{eq:poly_conv}
W_{m+1}(t) &= \left(\frac{d^{m+1}}{dt^{m+1}} + \sum_{\mu=0}^ml_\mu\frac{d^\mu}{dt^{\mu}}\right)U(t) \\
&=\left(\frac{d}{dt}+\gamma_m\right)\cdots\left(\frac{d}{dt}+\gamma_0\right)U(t)
\end{aligned}
\end{equation}
by the relation between the coefficients $l_0,\dots,l_m$ and $\gamma_0,\dots,\gamma_m$. Thus, define $V_0(t) = U_0(t)$ and recursivelly $V_{\mu+1}(t) = (d/dt + \gamma_\mu)V_\mu(t)$ for $0\leq\mu< m$ from which $W_{m+1}(t) = (d/dt+\gamma_m)V_{m}(t)$ is obtained using \eqref{eq:poly_conv}. Equivalently, we have
$
\dot{V}_\mu(t) = V_{\mu+1}(t) - \gamma_\mu V_\mu(t),\  0\leq\mu<m
$
and $\dot{V}_m(t) = W_{m+1}(t)-\gamma_mV_m(t)$. Written in vector form, $\mf{V}(t) := [V_0(t)^T,\dots,V_m(t)^T]^T$ satisfies $\dot{\mf{V}}(t) = (\Gamma\otimes \text{\normalfont I}_n)\mf{V}(t) + (B\otimes \text{\normalfont I}_n)W_{m+1}(t)$. Now, we obtain the matrix which maps $\mf{V}(t)$ to $\mf{U}(t)$ by noting that with $V_0(t) \equiv U_0(t)= (C\otimes \text{\normalfont I}_n)\mf{V}(t)$, 
$$
\begin{aligned}
U_1(t) = \dot{U}_0(t) &= (C\Gamma\otimes \text{\normalfont I}_n)\mf{V}(t) + (CB\otimes \text{\normalfont I}_n)W_{m+1}(t) \\&= (C\Gamma\otimes \text{\normalfont I}_n)\mf{V}(t)
\end{aligned}
$$
since $CB=0$ and continuing this procedure to obtain $U_\mu(t) = \dot{U}_{\mu-1}(t) = (C\Gamma^{\mu}\otimes \text{\normalfont I}_n)\mf{V}(t)$ since $C\Gamma^{\mu-1}B=0$ for $0\leq\mu<m$. This can be written as $\mf{U}(t) = (G\otimes \text{\normalfont I}_n)\mf{V}(t)$ or equivalently $\mf{V}(t) = (G^{-1}\otimes \text{\normalfont I}_n)\mf{U}(t) = \mf{W}(t)$. Therefore, $\mf{W}(t)$ satisfy \eqref{eq:dynW} concluding the proof.
\end{pf}
\begin{cor}
\label{cor:companion}
Let $\tilde{\Gamma},B$ be defined as in Lemma \ref{le:w_dynamics}. Then, $\tilde{\Gamma} = G\Gamma G^{-1}$ and $GB= B$.
\end{cor}
\vspace{-1em}
\begin{pf}
Let the change of variables from Lemma \ref{le:w_dynamics} as $\mf{U}(t) = (G\otimes \text{\normalfont I}_n)\mf{W}(t)$. Then,
$$
\begin{aligned}
\dot{\mf{U}} 
&= (G\otimes \text{\normalfont I}_n)(\Gamma\otimes \text{\normalfont I}_n)\mf{W} + (G\otimes \text{\normalfont I}_n)(B\otimes \text{\normalfont I}_n)W_{m+1} \\
&=(G\Gamma G^{-1}\otimes \text{\normalfont I}_n)\mf{U} + (GB\otimes \text{\normalfont I}_n)W_{m+1}\\ 
\end{aligned} 
$$
Comparing with \eqref{eq:dynU} completes the proof.
\end{pf}

\section{Convergence of the consensus components of REDCHO}
In this section we show the behaviour of $\bar{y}(t) = (\text{\normalfont I}_{m+1}\otimes \mathds{1}_n^T/n)\mf{Y}(t)$ as given in the following result.
\begin{lem}
\label{eq:conv_bar_y}
Let $\mf{Y}(t)\in\mathbb{R}^{n(m+1)}$. Then, with any $\gamma_0,\dots,\gamma_m>0$ and any initial conditions $\mf{Y}(t_0)$ for \eqref{eq:dyn_y}
it is satisfied that $\bar{y}(t)=(\text{\normalfont I}_{m+1}\otimes \mathds{1}_n^T/n)\mf{Y}(t)$ converge asymptotically towards $\mf{\bar{u}}(t)=(\text{\normalfont I}_{m+1}\otimes \mathds{1}_n^T/n)\mf{U}(t)=\left[\bar{u}(t)\mathds{1}_n^T,\dots,\bar{u}^{(m)}(t)\mathds{1}_n^T\right]^T$.
\end{lem}

\begin{pf}
First, recall that $G\Gamma G^{-1}=\tilde{\Gamma}$ from Corollary \ref{cor:companion} and obtain the dynamics of $\bar{y}(t)$ by multiplying \eqref{eq:dyn_y} by $(\text{\normalfont I}_{m+1}\otimes \mathds{1}_n^T/n)$ from the left:
$$
\begin{aligned}
\dot{\bar{y}} &= \mf{\dot{\bar{u}}} + (\text{\normalfont I}_{m+1}\otimes \mathds{1}_n^T/n)(\tilde{\Gamma}\otimes \text{\normalfont I}_n)(\mf{Y}-\mf{U})\\& + (\text{\normalfont I}_{m+1}\otimes \mathds{1}_n^T/n)(G\otimes \text{\normalfont I}_n)\mf{F}(Y_0;\theta)\\
&=\mf{\dot{\bar{u}}} + \tilde{\Gamma}(\bar{y}-\mf{\bar{u}}) +G(\text{\normalfont I}_{m+1}\otimes \mathds{1}_n^T/n)\mf{F}(Y_0;\theta)
\end{aligned}
$$
since $(\text{\normalfont I}_{m+1}\otimes \mathds{1}_n^T/n)(\tilde{\Gamma}\otimes \text{\normalfont I}_n)=(\tilde{\Gamma}\otimes\mathds{1}_n^T/n)=(\tilde{\Gamma}\otimes \text{\normalfont I}_1)(\text{\normalfont I}_{m+1}\otimes\mathds{1}_n^T/n) = \tilde{\Gamma}(\text{\normalfont I}_{m+1}\otimes \mathds{1}_n^T/n)$. Moreover, $G(\text{\normalfont I}_{m+1}\otimes \mathds{1}_n^T/n)\mf{F}(Y_0;\theta)=0$ since $\mathds{1}_n^TD=0$ \cite[Page 280]{Godsil}. Hence, we obtain $\dot{\bar{y}}(t) = \mf{\dot{\bar{u}}}(t) + \tilde{\Gamma}(\bar{y}(t) - \bar{\mf{u}})(t)$. Define the error $e(t) := \bar{y}(t) - \bar{\mf{u}}(t)$ to obtain $\dot{e}(t)=\tilde{\Gamma}e(t)$. Finally, note that $\tilde{\Gamma}$ is given in \eqref{eq:companion} and has characteristic polynomial $\prod_{\mu=0}^m(s+\gamma_\mu)=0$. Then, with $\gamma_0,\dots,\gamma_m>0$, $\tilde{\Gamma}$ has negative eigenvalues and $\bar{y}(t)-\mf{\bar{u}}(t)$ asymptotically converge to the origin.
\end{pf}

\section{Convergence of the consensus error}
In this section we show the behaviour of $\tilde{\mf{Y}}(t) = (\text{\normalfont I}_{m+1}\otimes P)\mf{Y}(t)$ where $P = (\text{\normalfont I}_n-(1/n)\mathds{1}_n\mathds{1}_n^T)$. First, we obtain how the dynamics of $\tilde{\mf{Y}}(t)$ relate to EDCHO.
\begin{lem}
Let  $\tilde{\mf{Y}}(t) = (\text{\normalfont I}_{m+1}\otimes P)\mf{Y}(t)$ where $P = (\text{\normalfont I}_n-(1/n)\mathds{1}_n\mathds{1}_n^T)$ and $\mf{Y}(t)$ satisfy \eqref{eq:dyn_y}. Moreover, let $\mf{W}(t) :=[W_0^T,\dots,W_m^T]^T= (G^{-1}\otimes \text{\normalfont I}_n)\mf{U}(t)$ with $W_\mu(t)\in\mathbb{R}^n, \forall\mu\in\{0,\dots,m\}$ \RevFour{ and $\Theta=\diag{1,\theta^{-1},\dots,\theta^{-m}}$. Then, $\mf{Z}(t) := [Z_0(t)^T,\dots,Z_m(t)^T]^T = (\Theta G^{-1}\otimes \text{\normalfont I}_n)\mf{\tilde{Y}}(t)$ with $Z_\mu(t)\in\mathbb{R}^n$, satisfy}
\RevFour{
\begin{equation}
\label{eq:dynZ}
    \begin{aligned}
        \dot{Z}_{\mu}(t) = \theta\Big(&Z_{\mu+1}(t) - k_\mu D\sgn{D^T Z_{0}(t)}{\frac{m-\mu}{m+1}}\\& - (\gamma_\mu/\theta) Z_\mu(t)\Big) \text{ for } 0\leq\mu \leq m-1, \\
    \dot{Z}_{m}(t) =\theta\Big(&PW_{m+1}(t)/\theta^{m+1}- k_mD\sgn{D^TZ_0(t)}{0}\\& - (\gamma_m/\theta)Z_m(t)\Big)
    \end{aligned}
\end{equation}
}
with $W_{m+1}(t)$ defined in \eqref{eq:Wmp1}.
\end{lem}

\begin{pf}
First, obtain the dynamics of $\tilde{\mf{Y}}(t)$ using \eqref{eq:dyn_y} and $\tilde{\Gamma}=G\Gamma G^{-1}$ from Corollary \ref{cor:companion}:
$$
\begin{aligned}
\frac{\text{d}}{\text{d}t}{\mf{\tilde{Y}}} &= (\text{\normalfont I}_{m+1}\otimes P)\dot{\mf{U}} + (\tilde{\Gamma}\otimes \text{\normalfont I}_n)\mf{\tilde{Y}} \\&- (\tilde{\Gamma}\otimes P)\mf{U} - (G\otimes P)\mf{F}(Y_0;\theta)
\end{aligned}
$$
However, note that $(\text{\normalfont I}_{m+1}\otimes P)\dot{\mf{U}}(t) = (\tilde{\Gamma}\otimes P)\mf{U}(t) + (B\otimes P)W_{m+1}(t)$ from \eqref{eq:dynU} in Lemma \ref{le:w_dynamics}. Then,
$$
    \frac{\text{d}}{\text{d}t}{\mf{\tilde{Y}}} = (\tilde{\Gamma}\otimes \text{\normalfont I}_n)\tilde{\mf{Y}} - (G\otimes P)\mf{F}(Y_0;\theta) + (B\otimes P)W_{m+1}
$$
Moreover, $(G\otimes P)\mf{F}(Y_0(t);\theta) = (G\otimes \text{\normalfont I}_n)\mf{F}(Y_0(t);\theta)$ since
\RevFour{
} $PD=(\text{\normalfont I}_n-(1/n)\mathds{1}_n\mathds{1}_n^T)D = D$. Furthermore, the dynamics of $\mf{Z}(t)$ are
\RevFour{$$
\begin{aligned}
\dot{\mf{Z}} &= (\Theta G^{-1}\otimes \text{\normalfont I}_n) \frac{\text{d}}{\text{d}t}\tilde{\mf{Y}} \\
&=(\Theta G^{-1}\tilde{\Gamma}G\Theta^{-1}\otimes \text{\normalfont I}_n)\mf{Z} \\
&- (\Theta\otimes \text{\normalfont I}_n)\mf{F}(Y_0;\theta) + (\Theta G^{-1}B\otimes P)W_{m+1} \\
&=(\Theta\Gamma\Theta^{-1}\otimes \text{\normalfont I}_n)\mf{Z}\\& - (\Theta\otimes \text{\normalfont I}_n)\mf{F}(Y_0;\theta) + \theta^{-m}(B\otimes \text{\normalfont I}_n)PW_{m+1}
\end{aligned}
$$
where Corollary \ref{cor:companion} was used and $(\Theta B\otimes P) = \theta^{-m}(B\otimes \text{\normalfont I}_n)(\text{\normalfont I}_1\otimes P) = \theta^{-m}(B\otimes \text{\normalfont I}_n)P$. In addition, note that}
\RevFour{
$$
\begin{aligned}
&(\Theta\otimes \text{\normalfont I}_{n})\mf{F}(Y_0;\theta)\\& = \begin{bmatrix}
\text{\normalfont I}_n & 0 & \cdots & 0 \\
0 & \text{\normalfont I}_n\theta^{-1} & \ddots & \vdots \\
\vdots& \ddots & \ddots & 0\\
0 &\cdots & 0& \text{\normalfont I}_n\theta^{-m}
\end{bmatrix}
\begin{bmatrix}
 k_0\theta D\sgn{D^TY_{0}}{\frac{m}{m+1}}\\
\vdots\\
k_m\theta^{m+1}D\sgn{D^TY_{0}}{0}
\end{bmatrix} \\&= \theta\mf{F}(Y_0;1)
\end{aligned}
$$
To simplify the $\Theta\Gamma\Theta^{-1}$ term, denote the nilpotent matrix $A_0:=\Gamma+\diag{\gamma}$ which does not depend on any of the parameters $\gamma=[\gamma_0,\dots,\gamma_{m}]^T$ and for which it can be verified $\Theta A_0 = \theta A_0 \Theta$. Hence,  
$$
\begin{aligned}
&\Theta\Gamma\Theta^{-1}=(\Theta A_0-\Theta\diag{\gamma})\Theta^{-1} \\
&=(\theta A_0\Theta-\diag{\gamma}\Theta ) \Theta^{-1}= \theta (A_0-\diag{\gamma/\theta})=\theta\Gamma_\theta
\end{aligned}
$$
Where $\Gamma_\theta:=A_0-\diag{\gamma/\theta}$. Moreover, the first row of $G\Theta^{-1}$ is $C$ which leads to $Z_0(t)=Y_0(t)$. Hence, combining all these facts,}
\RevFour{
$$
\begin{aligned}
&\dot{\mf{Z}} =  (\theta\Gamma_\theta\otimes \text{\normalfont I}_n)\mf{Z} - \theta\mf{F}(Z_0;1) + \theta^{-m}(B\otimes \text{\normalfont I}_n)PW_{m+1}\\
&=\theta\left((\Gamma_\theta\otimes \text{\normalfont I}_n)\mf{Z} - \mf{F}(Z_0;1) + \theta^{-(m+1)}(B\otimes \text{\normalfont I}_n)PW_{m+1}\right)
\end{aligned}
$$}
Writing this equation in partially vertorized form we recover \eqref{eq:dynZ}, which completes the proof.
\end{pf}
Now, if we show that $\mf{Z}(t)$ converge to the origin in finite time, the same conclusion will apply to $\tilde{\mf{Y}}(t)$. \RevFour{Note that for given $\gamma_0,\dots,\gamma_m$,  $PW_{m+1}(t)/\theta^{m+1}\in[-L,L]^n/\theta^{m+1}\subseteq[-L,L]^n$ under Assumption \ref{as:u_signals} and $\theta\geq 1$}. Therefore, comparing \eqref{eq:dynZ} with the EDCHO error system \eqref{eq:main_algo_error2} in Appendix \ref{ap:edcho}, it would be the case that $\mf{Z}(t)$ reaches the origin if $\gamma_0=\cdots=\gamma_m=0$. In the following we will use homogeneity to show that even with those terms, stability of REDCHO will still be valid locally. To do so, we will decompose the right hand side of \eqref{eq:dynZ} in two parts, one similar to the right hand side of the EDCHO error system in \eqref{eq:main_algo_error2} and the other with the remaining linear terms. Let $\mf{H}:\mathbb{R}^{n(m+1)}\rightrightarrows\mathbb{R}^{n(m+1)}$ and $\mf{Q}:\mathbb{R}^{n(m+1)}\to\mathbb{R}^{n(m+1)}$ be defined as
\begin{equation}
\label{eq:HG}
    \mf{H}(\mf{Z}) = \begin{bmatrix}
Z_1 \\
\vdots\\
Z_m\\
[-L,L]^n
\end{bmatrix} - \mf{F}(Z_0;1), \ \ \RevFour{ \mf{Q}(\mf{Z}) = \begin{bmatrix}
-(\gamma_0/\theta)Z_0\\
\vdots\\
-(\gamma_m/\theta)Z_m
\end{bmatrix}}
\end{equation}
Then, \eqref{eq:dynZ} is equivalent to the differential inclusion \RevFour{$\dot{\mf{Z}}(t) \in \theta(\mf{H}(\mf{Z}(t)) + \mf{Q}(\mf{Z}(t)))$}. Moreover, let  $\mf{r} = [r_0\mathds{1}^T,\dots,r_m\mathds{1}^T]$ with $r_\mu := m+1-\mu, \forall \mu\in\{0,\dots,m\}$. Then, it can be verified that $\mf{H}(\bullet)$ and $\mf{Q}(\bullet)$ are $\mathbf{r}$-homogeneous of degrees $-1$ and $0$ respectively in the sense of   in the sense of Definition \ref{def:homo} in Appendix \ref{sec:homo}. 

\begin{lem}
\label{le:z_convergence}
\RevFour{Let  $\mathcal{G}$ be a connected graph} and the pair $\mf{H}(\bullet)$, $\mf{Q}(\bullet)$ defined in \eqref{eq:HG}. Moreover, let some \RevFour{fixed $\gamma_0,\dots,\gamma_m>0$ so that there exists $L$ for which Assumption \ref{as:u_signals} is complied and $k_0,\dots,k_m$ chosen as in Theorem \ref{prop:edcho} in Appendix \ref{ap:edcho} for such $L$}. Then, there exists a neighborhood $\mathcal{R}_0\subset \mathbb{R}^{n(m+1)}$ of the origin such  that if $\mf{Z}(t_0)\in\mathcal{R}_0$, then the solution of \RevFour{$\dot{\mf{Z}}(t) \in \theta(\mf{H}(\mf{Z}(t))+ \mf{Q}(\mf{Z}(t)))$} converge to the origin in finite-time. Moreover, $\mathcal{R}_0$ can be made arbitrarily big by \RevFour{increasing $\theta$}.
\end{lem}
\begin{pf}
Consider $\gamma_0=\cdots=\gamma_m=0, \RevFour{\theta=1}$. Then, $\dot{\mf{Z}} \in \mf{H}(\mf{Z})$ is globally finite-time stable towards the origin by Theorem \ref{prop:edcho} from Appendix \ref{ap:edcho}. Moreover, recall that $\mf{H}(\bullet)$ is $\mf{r}$-homogeneous of degree $-1$. Hence, by Proposition \ref{prop:converse} in Appendix \ref{sec:homo} there exists scalar functions $V(\mf{Z}),W(\mf{Z})$ which are $\mf{r}$-homogeneous of degrees $k$ and $k-1$ respectively and comply $\dot{V}(\mf{Z}) = L_{\mf{H}}V(\mf{Z}) \leq -W(\mf{Z})\leq -\beta_mV(\mf{Z})^\frac{k-1}{k}$ with $0<\beta_m=\inf\{W(\mf{Z}):V(\mf{Z})=1\}$ using Proposition \ref{prop:homo_comparison}. Now consider $\gamma_0,\dots,\gamma_m>0$ with arbitrary $\theta\geq 1$ and the same Lyapunov function $V(\mf{Z})$ as before. In this case 
\RevFour{
$$
\dot{V} = \theta\left(L_{\mf{H}}V + L_{\mf{Q}}V\right) \leq  \theta\left(-\beta_mV^{\frac{k-1}{k}} + L_{\mf{Q}}V\right)
$$}
Note that from Proposition  \ref{prop:lie_homo}  $L_{\mf{Q}}V(\mf{Z})=\frac{\partial V}{\partial \mf{Z}}\mf{Q}(\mf{Z})$ is $\mf{r}$-homogeneous of degree $k$ since $\mf{Q}(\bullet)$ is $\mf{r}$-homogeneous of degree $0$. Hence, $L_{\mf{Q}}V(\mf{Z})\leq  \beta_M V(\mf{Z})$ by Proposition \ref{prop:homo_comparison} and $\beta_M=\sup\{L_{\mf{Q}}V(\mf{Z}):V(\mf{Z})=1\}$. Note that $L_{\mf{Q}}V(\mf{Z})$ may be positive for some $\mf{Z}$. Therefore, $\beta_M$ may be positive too. Moreover, 
\RevFour{
$$
\dot{V}\leq  \theta\left[-\beta_mV^{\frac{k-1}{k}} + \beta_MV\right]\leq -\theta\left[\beta_m - \beta_M V^{\frac{k}{k-1}} \right]V^{\frac{k-1}{k}}
$$
Denote with $\mathcal{R}_0\subseteq\mathbb{R}^{n(m+1)}$ any region in which 
\begin{equation}
\label{eq:condition}
-\left[\beta_m - \beta_M V(\mf{Z})^{\frac{k}{k-1}}\right]\leq -c
\end{equation}}
\RevFour{is complied for some $c>0$ so that $\dot{V}(\mf{Z})\leq -\theta cV(\mf{Z})^{\frac{k-1}{k}}$ for any $\mf{Z}\in\mathcal{R}_0$. If $\beta_M\leq 0$, then \eqref{eq:condition} is complied for $c=\beta_m$ regardless of $\mf{Z}$ so that we can set $\mathcal{R}_0=\mathbb{R}^{n(m+1)}$. On the other hand, if $\beta_M>0$ then \eqref{eq:condition} is complied when 
$
V(\mf{Z})^{\frac{k}{k-1}} \leq (\beta_m-c)/\beta_M
$ which is possible only for $c\in(0,\beta_m)$. Then, choose $\mathcal{R}_0$ with $c=\beta_m/2$ so that \eqref{eq:condition} is complied whenever $V(\mf{Z})\leq \beta_m/(2\beta_M)$. We can write explicitly $\mathcal{R}_0=\{\mf{Z}\in\mathbb{R}^{n(m+1)}:V(\mf{Z})\leq \beta_m/(2\beta_M)\}$ so that $\dot{V}(\mf{Z})\leq -(\theta\beta_m/2)V(\mf{Z})^{\frac{k-1}{k}}$ for any $\mf{Z}\in\mathcal{R}_0$ regardless of $\beta_M$. 
}
\RevFour{
Note that due to Proposition \ref{prop:converse}, $V(\mf{Z})$ is continuously differentiable and we can write $\beta_M = \sup\left\{-\sum_{\mu=0}^m(\gamma_\mu/\theta)\frac{\partial V}{\partial Z_\mu}Z_\mu: V(\mf{Z})=1\right\} = \tilde{\beta}_M/\theta$ where $\tilde{\beta}_M:=\sup\left\{-\sum_{\mu=0}^m\gamma_\mu\frac{\partial V}{\partial Z_\mu}Z_\mu: V(\mf{Z})=1\right\}$ is a constant for fixed $\gamma_0,\dots,\gamma_m$. Thus, $\beta_m/(2\beta_M)=\theta (\beta_m/(2\tilde{\beta}_M))$ can be made arbitrarily big by increasing $\theta$, so that $\mathcal{R}_0$ can be made arbitrarily big as well. Finally, note that since $(k-1)/k\in(0,1)$, then $V(\mf{Z})$ will reach the origin in finite time \cite[Corolary 4.25]{bernuau2014} and so will $\mf{Z}(t)$ whenever $\mf{Z}(t_0)\in\mathcal{R}_0$, completing the proof. 
}
\end{pf}

The previous result shows that trajectories of $\dot{\mf{Z}}(t)\in\theta(\mf{H}(\mf{Z}(t))+\mf{Q}(\mf{Z}(t)))$ reach the origin if $\mf{Z}(t_0)\in\mathcal{R}_0$ which motivates to study if diverging trajectories can be obtained for some $\mf{Z}(t_0)\notin\mathcal{R}_0$. In the following, we show that this is not possible and only a terminal bounded error is allowed.

\begin{lem}
\label{le:bounded_error}
Let the conditions of Lemma \ref{le:z_convergence} be satisfied. Thus, for any initial conditions $\mf{Z}(t_0)\in\mathbb{R}^{n(m+1)}$, there exists $T>0$ and a bounded neighborhood of the origin $\mathcal{R}_\infty$ such that solution of \RevFour{$\dot{\mf{Z}}(t) \in\theta( \mf{H}(\mf{Z}(t))+\mf{Q}(\mf{Z}(t)))$} comply $\mf{Z}(t)\in\mathcal{R}_\infty$ for $t\geq T+t_0$. Moreover, such neighborhood can be made arbitrarily big \RevFour{by increasing $\theta$}.
\end{lem}
\begin{pf}
We proceed very similarly to the proof of Lemma \ref{le:z_convergence}. Consider only $\dot{\mf{Z}}(t)=\theta\mf{Q}(\mf{Z}(t))$ using \eqref{eq:HG} and a Lyapunov function $V(\mf{Z}) = \sum_{\mu=0}^m({Z}_\mu^T{Z}_\mu)^{\frac{m+1}{2r_\mu}}$ \RevFour{ with $r_\mu := m+1-\mu, \forall \mu\in\{0,\dots,m\}$} obtaining
\RevTwo{
$$
\dot{V} = \sum_{\mu=0}^m\frac{m+1}{r_\mu}({Z}_\mu^T{Z}_\mu)^{\frac{m+1}{2r_\mu}-1}{Z}_\mu^T\left({-\gamma_\mu} {Z}_\mu\right)\leq {-\gamma_{\min}}V 
$$
since $\gamma_{\min}:=\min\{\gamma_0,\dots,\gamma_m\}\leq \gamma_\mu$ and $1\leq\frac{m+1}{r_\mu}$, $\forall \mu\in\{0,\dots,m\}$.} Moreover, note that 
$
V(\Delta_\mf{r}(\lambda)\mf{Z}) = \sum_{\mu=0}^m(\lambda^{2r_\mu}Z_\mu^TZ_\mu)^{\frac{m+1}{2r_\mu}} = \lambda^{m+1}V(\mf{Z})
$
and thus $V(\mf{Z})$ is $\mf{r}$-homogenenous of degree $m+1$ under the dilation $\Delta_{\mf{r}}(\lambda)\mf{Z} = [\lambda^{r_0}Z_0^T,\dots,\lambda^{r_m}Z_m^T]^T$. Let the same Lyapunov function for \RevFour{$\dot{\mf{Z}}(t) \in \theta( \mf{H}(\mf{Z}(t))+\mf{Q}(\mf{Z}(t)))$:
$$
\dot{V}= \theta\left(L_{\mf{Q}}V+L_{\mf{H}}V\right)\leq-{\gamma_{\min}}V + \theta L_{\mf{H}}V.
$$}
Note that from Proposition  \ref{prop:lie_homo}  $L_{\mf{H}}V(\mf{Z})$ is $\mf{r}$-homogeneous of degree $m$ since $\mf{H}(\bullet)$ is $\mf{r}$-homogeneous of degree $-1$. Hence, $L_{\mf{H}}V(\mf{Z})\leq  \beta_M' V(\mf{Z})^\frac{m}{m+1}$ by Proposition \ref{prop:homo_comparison} and $\beta_M'=\sup\{L_{\mf{H}}V(\mf{Z}):V(\mf{Z})=1\}$. Thus,
\RevFour{
$$
\dot{V}\leq-\left[{\gamma_{\min}} -\theta\beta_M' V^{-\frac{1}{m+1}}\right]V.
$$}
\RevFour{
Denote with $\mathcal{R}_\infty^c\subset\mathbb{R}^{n(m+1)}$ any region in which 
\begin{equation}
    \label{eq:condition2}
    -\theta\left[\frac{\gamma_{\min}}{\theta} -\beta_M' V^{-\frac{1}{m+1}}\right] \leq -\theta c'
\end{equation}
for some $c'>0$ so that $\dot{V}(\mf{Z})\leq -\theta c'V(\mf{Z})$ for any $\mf{Z}\in\mathcal{R}_\infty^c$. If $\beta_M'\leq 0$, then we can set $\mathcal{R}_\infty^c=\mathbb{R}^{n(m+1)}$ with $c'=\gamma_{\min}/\theta$. On the other hand if $\beta_M'>0$, \eqref{eq:condition2} is equivalent to
$
V(\mf{Z})^{1/(m+1)}\geq \beta_M'/(\gamma_{\min}/\theta-c')
$. Then, choose $c'=\gamma_{\min}/(2\theta)$ so that we can write $\mathcal{R}_\infty^c=\{\mf{Z}\in\mathbb{R}^{n(m+1)} : V(\mf{Z})^{1/k}\geq 2\theta\beta_M'/\gamma_{\min}\}$ so that $\dot{V}(\mf{Z})\leq -(\gamma_{\min}/2)V(\mf{Z})$ for any $\mf{Z}\in\mathcal{R}_{\infty}^c$. Therefore, for any initial condition $\mf{Z}(t_0)\in\mathbb{R}^{n(m+1)}$ the trajectory of ${V(\mf{Z}(t))}$ will converge to $\mathcal{R}_\infty:=\mathbb{R}^{n(m+1)}\setminus\mathcal{R}_\infty^c$ after a finite time $T+t_0$ and comply $\mf{Z}(t)\in\mathcal{R}_\infty$ for all $t\geq T+t_0$. Finally, note that $\mathcal{R}_\infty$ can be made arbitrarily large by increasing $\theta$.}

\end{pf}
\section{Convergence of REDCHO}
\label{sec:main_result}
In this section we formally state the main result of this work.
\begin{thm} 
\label{th:main}
\RevFour{
\RevFour{Let  $\mathcal{G}$ be a connected graph} and the pair $\mf{H}(\bullet)$, $\mf{Q}(\bullet)$ defined in \eqref{eq:HG}. Moreover, let some \RevFour{fixed $\gamma_0,\dots,\gamma_m>0$ so that there exists $L$ for which Assumption \ref{as:u_signals} is complied and $k_0,\dots,k_m$ chosen as in Theorem \ref{prop:edcho} in Appendix \ref{ap:edcho} for such $L$}. Then, there exists neighborhoods $\mathcal{R},\mathcal{R}'\subset \mathbb{R}^{n(m+1)}$ around consensus such that if the initial conditions comply $[y_{1,0}(t_0),\dots,y_{n,m}(t_0)]^T\in\mathcal{R}$, the REDCHO algorithm in \eqref{eq:redcho} achieves robust EDC. \RevEight{On the other hand, if $[y_{1,0}(t_0),\dots,y_{n,m}(t_0)]^T\notin\mathcal{R}$, \eqref{eq:redcho} will achieve at most a uniformly bounded terminal error $[y_{1,0}(t),\dots,y_{n,m}(t)]^T\in\mathcal{R}',\forall t\geq T$ around dynamic consensus after some finite time $T>0$. Moreover, the neighborhoods $\mathcal{R},\mathcal{R}'$ can be made arbitrarily big by increasing $\theta\geq 1$.}}
\end{thm}
\begin{pf}
First, decompose $\mf{Y}(t) = (I\otimes \mathds{1}_n)(\text{\normalfont I}_{m+1}\otimes\mathds{1}_n^T/n)\mf{Y}(t) + (\text{\normalfont I}_{m+1}\otimes P)\mf{Y}(t) = (\text{\normalfont I}_{m+1}\otimes \mathds{1})\bar{y}(t) + \tilde{\mf{Y}}(t)$. \RevTwo{Now, note that Lemma \ref{le:z_convergence} implies the existence of a neighborhood $\mathcal{R}_0$ such that if $\mf{Z}(t_0)\in\mathcal{R}_0$ with $\mf{Z}(t) := (\Theta G^{-1}\otimes \text{\normalfont I}_n)\tilde{\mf{Y}}(t)$ then convergence of $\mf{Z}(t)$ is achieved towards the origin. Hence, consider the biggest ball of radius $R(\theta)$, $\mathcal{B}_0(\theta)=\{\mf{Z}\in\mathbb{R}^{n(m+1)}:\mf{Z}^T\mf{Z}\leq R(\theta)^2\}$ such that $\mathcal{B}_0(\theta)\subseteq\mathcal{R}_0$ and $R(\theta)$ is an increasing function of $\theta$ due to the last part of Lemma \ref{le:z_convergence}. Now, 
$$
\begin{aligned}
R(\theta)^2&\geq \mf{Z}^T\mf{Z}=\mf{Y}^T(\Theta (G^{-1})^TG^{-1}\Theta\otimes \text{\normalfont I}_n)\mf{Y}\\
&\geq \theta^{-2m}\mf{Y}^T((G^{-1})^TG^{-1}\otimes \text{\normalfont I}_n)\mf{Y}
\end{aligned}
$$
This implies that convergence of $\tilde{\mf{Y}}(t)$ towards the origin happens for any $\mf{\tilde{Y}}(t_0)\in\{\mf{Y}\in\mathbb{R}^{n(m+1)}:\mf{Y}^T((G^{-1})^TG^{-1}\otimes \text{\normalfont I}_n)\mf{Y}\leq (\theta^{m}R(\theta))^2 \}$ where the previous region can be made arbitrarily big by increasing $\theta$. This implies the existence of $\mathcal{R}\in\mathbb{R}^{n(m+1)}$ as required by the theorem. An identical argument can be made for $\mathcal{R}'\in\mathbb{R}^{n(m+1)}$ but using Lemma \ref{le:bounded_error} instead to conclude uniformly bounded trajectories for $\tilde{\mf{Y}}(t)$ implying uniformly bounded steady state error around dynamic consensus.} Furthermore, Lemma \ref{eq:conv_bar_y} imply that $\bar{y}(t)$ converge asymptotically towards $\mf{\bar{u}}(t)=(\text{\normalfont I}_{m+1}\otimes \mathds{1}_n^T/n)\mf{U}(t)=\left[\bar{u}(t)\mathds{1}_n^T,\dots,\bar{u}^{(m)}(t)\mathds{1}_n^T\right]^T$. Hence, when $\mf{Y}(t_0)\in\mathcal{R}$, then $\mf{Y}(t)$ converge asymptotically  towards $(\text{\normalfont I}_{m+1}\otimes \mathds{1}_n)\mf{\bar{u}}(t)$. Equivalently, $Y_\mu(t)\to \bar{u}^{(\mu)}(t)\mathds{1}_n$. Since no initialization condition is required, then \eqref{eq:redcho} achieves robust EDC. 
\end{pf}
\vspace{-1cm}
\RevFour{
\begin{rem}
\label{rem:params}
Note that since the $\gamma_0,\dots,\gamma_m$ are fixed, the class of signals for which Assumption \ref{as:u_signals} is complied can be checked before-hand, so that the method remains fully distributed.
On the other hand, showing the same stability properties in the case when all agents have different parameters $\gamma_0^i,\dots,\gamma_m^i>0, i\in\{1,\dots,n\}$ require more complicated computations, but is straightforward using similar arguments as in this work. Thus, only the case with a single set of parameters for all agents is provided here for simplicity.    
\end{rem}}

\section{Simulation examples}
\label{sec:simulations}
In the following we show some simulation scenarios designed to show the properties of the REDCHO protocol. The simulations were  implemented using explicit Euler method with time step $h=10^{-6}$ over \eqref{eq:redcho}.

\begin{figure}
    \centering
    \includegraphics[width=0.2\textwidth]{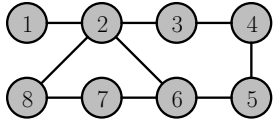}
    \caption{The network topology $\mathcal{G}$ considered in Example \ref{ex:convergence}.}
    \label{fig:g1}
\end{figure}

\begin{figure}
    \centering
    \includegraphics[width=0.45\textwidth]{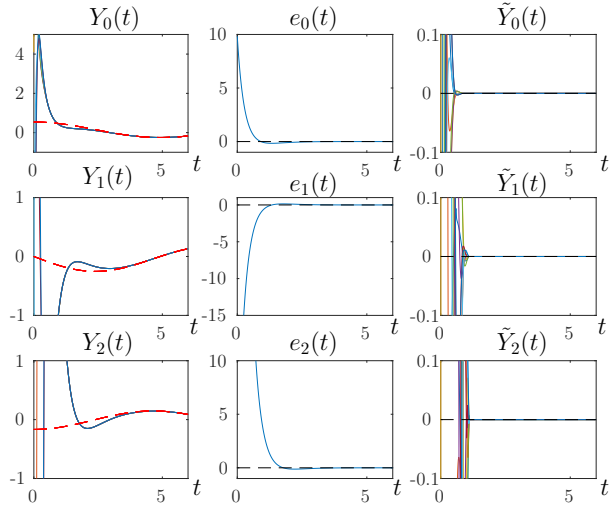}
    \caption{Convergence of the REDCHO algorithm for the scenario in Example \ref{ex:convergence}. The first column shows convergence of ${Y}_0(t),{Y}_1(t),{Y}_2(t)$ towards the signals $\bar{u}(t)\mathds{1},\dot{\bar{u}}(t)\mathds{1},\ddot{\bar{u}}(t)\mathds{1}$. The second column shows the average consensus error $e_\mu(t):=\bar{y}_\mu(t)-\bar{u}^{(\mu)}(t), \mu=0,1,2$. The third column shows the consensus errors $\tilde{Y}_0(t),\tilde{Y}_1(t),\tilde{Y}_2(t)$.}
    \label{fig:convergence}
\end{figure}

\begin{figure}
    \centering
    \includegraphics[width=0.45\textwidth]{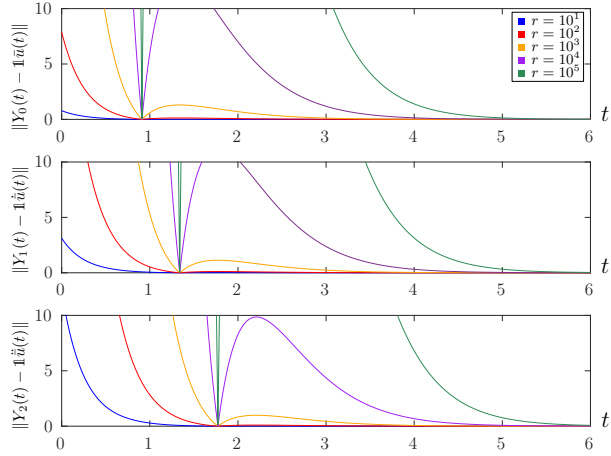}
    \caption{Convergence of the REDCHO algorithm for different magnitudes of initial conditions as described in Example \ref{ex:convergence}.}
    \label{fig:initial}
\end{figure}

\begin{example}
\label{ex:convergence}
In order to show the convergence properties as described in the analysis from the previous sections, we simulated \eqref{eq:redcho} for the network topology $\mathcal{G}$ shown in Figure \ref{fig:g1}. Moreover,the number of agents is $n=8$ and the signals $u_i(t)=a_i\cos(\omega_it)$ with amplitudes $a_i=0.95, 0.34, 0.58, 0.22, 0.75, 0.25, 0.50, 0.69$ and frequencies $\omega_i=0.70, 0.75, 0.27, 0.67, 0.65, 0.16, 0.11, 0.49$. Thus, we choose $m=2$, $\gamma_0=\gamma_1=\gamma_2=3$, $k_0=6, k_1=11, k_2=6, \theta=1.5$. Furthermore, initial conditions for \eqref{eq:redcho} were generated from a normal distribution with mean $1$ and variance $r=1$. Figure \ref{fig:convergence} shows the convergence of ${Y}_0(t),{Y}_1(t),{Y}_2(t)$ towards the signals $\bar{u}(t)\mathds{1},\dot{\bar{u}}(t)\mathds{1},\ddot{\bar{u}}(t)\mathds{1}$ in the first column. Moreover, by letting $\bar{y}(t)=[\bar{y}_0(t),\dots,\bar{y}_m(t)]^T$, we show the average consensus error $e_\mu(t):=\bar{y}_\mu(t)-\bar{u}^{(\mu)}(t), \mu=0,1,2$ in the second column of Figure \ref{fig:convergence}, \RevTwo{which converges asymptotically to the origin as expected from Lemma \ref{eq:conv_bar_y}.} The third column of Figure \ref{fig:convergence} \RevTwo{shows how the consensus errors $\tilde{Y}_0(t),\tilde{Y}_1(t),\tilde{Y}_2(t)$ converge to the origin in finite time as expected from Lemma \ref{le:z_convergence}.} This same experiment was repeated for different values of $r$ for which the norm of the average consensus errors $\|Y_\mu(t)-\bar{u}^{(\mu)}(t)\mathds{1}\|, \mu=0,1,2$ is shown in Figure \ref{fig:initial}. This experiment shows that even for big initial conditions, the algorithm manages to converge to EDC. \RevTwo{This is consistent with Lemma \ref{le:bounded_error} which implies that no diverging error trajectories are possible.}
\end{example}

\begin{figure}
    \centering
    \includegraphics[width=0.25\textwidth]{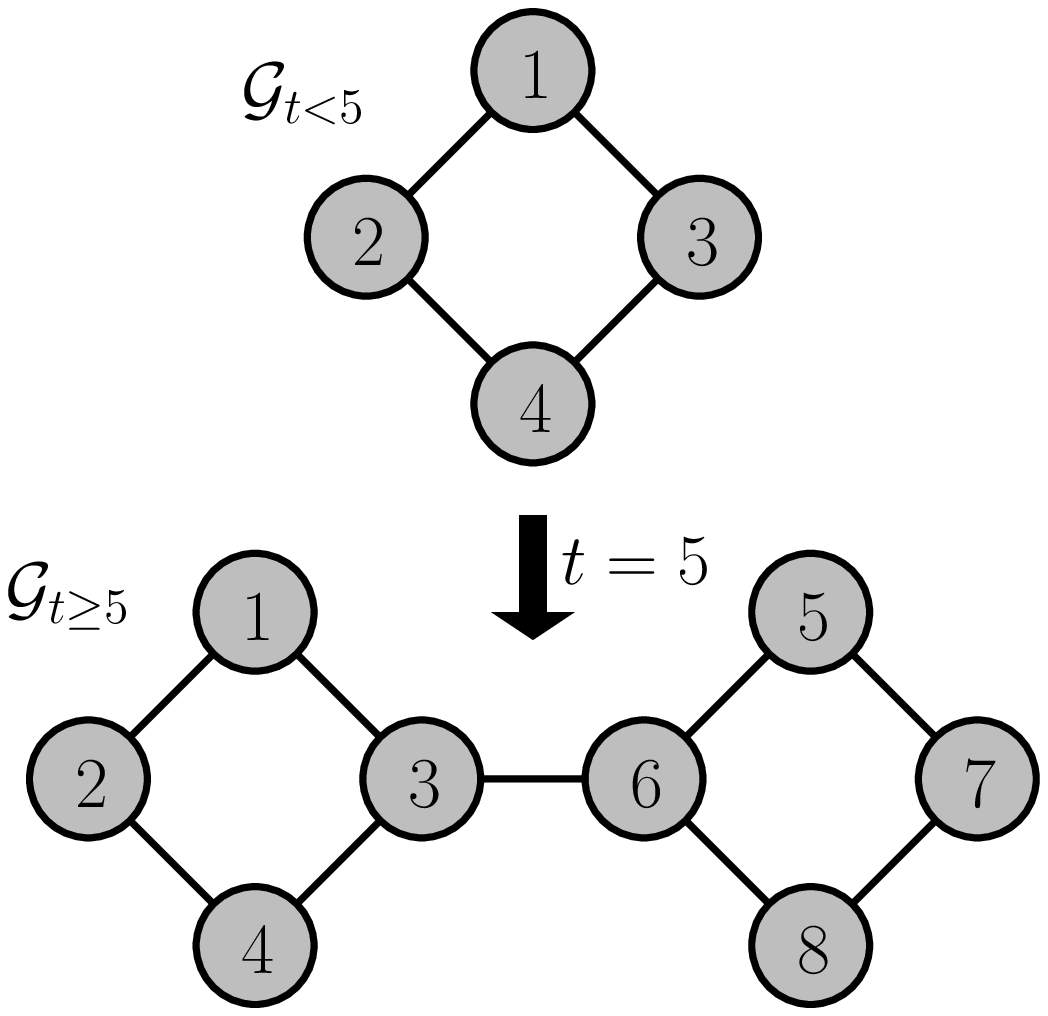}
    \caption{The network $\mathcal{G}$ considered in Example \ref{ex:merge}.}
    \label{fig:merge_graph}
\end{figure}

\begin{figure}
    \centering
    \includegraphics[width=0.45\textwidth]{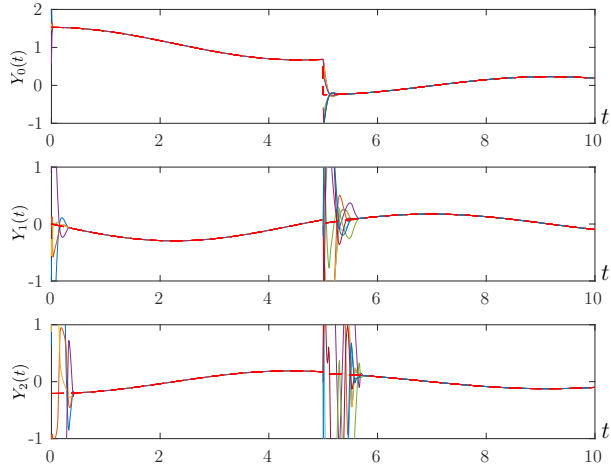}
    \caption{Trajectories for the REDCHO protocol in the scenario of Figure \ref{fig:merge_graph}, as presented in Example \ref{ex:merge}.}
    \label{fig:merge_trajectories}
\end{figure}

\begin{figure}
    \centering
    \includegraphics[width=0.45\textwidth]{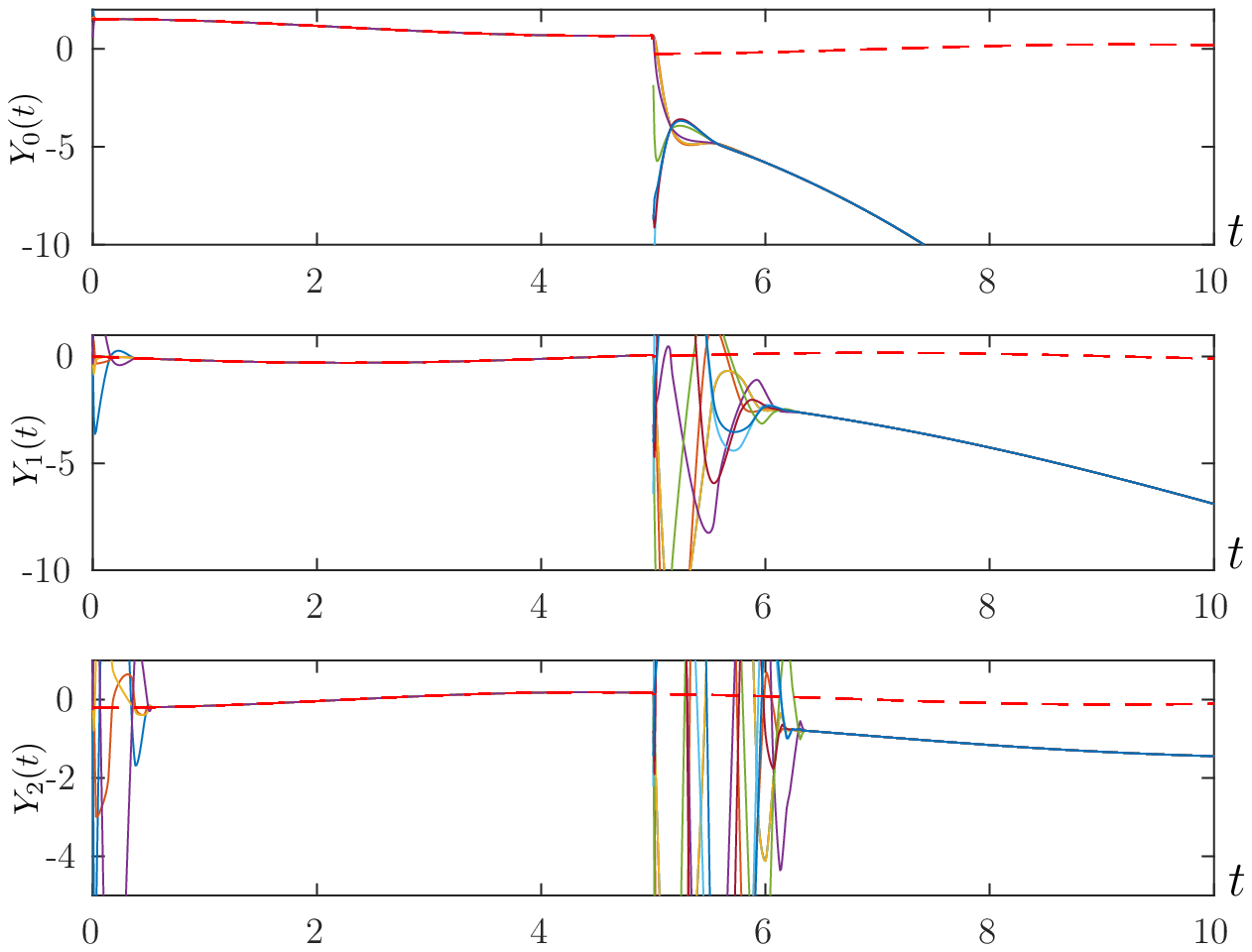}
    \caption{Trajectories for the protocol from \cite{aldana2021} scenario of Figure \ref{fig:merge_graph}, as presented in Example \ref{ex:merge}.}
    \label{fig:merge_trajectories_edcho}
\end{figure}

\begin{example}
\label{ex:merge}
In order to show the robustness properties of the REDCHO protocol when the topology suffers from sudden changes, we simulated \eqref{eq:redcho} for the network topology shown in Figure \ref{fig:merge_graph} which changes from $\mathcal{G}_{t< 5}$ to $\mathcal{G}_{t\geq 5}$ at $t=5$. Consider the same configuration and signals as in the previous example. Figure \ref{fig:merge_trajectories} shows how the outputs of the REDCHO algorithm converge to EDC approximately at $t=0.5$. At $t=5$ the topology changes, but the REDCHO protocol manages to make all agents, the first four and the new ones, converge to EDC again even when the states didn't comply neither $\sum_{i=1}^4x_{i,\mu}(0)=0$ nor $\sum_{i=1}^8x_{i,\mu}(5)=0$  as required in \cite{aldana2021}. For comparison, consider the protocol in \cite{aldana2021} obtained by setting $\gamma_0=\gamma_1=\gamma_2=0$ in REDCHO. Moreover, initial conditions are changed so that $\sum_{i=1}^4x_{i,\mu}(0)=0$. Figure \ref{fig:merge_trajectories_edcho} shows the trajectories for the protocol in this case, where EDC is achieved before $t=5$. However, when the new agents merge to the network, the agents output converge to consensus towards a signal that diverges.
\end{example}


\RevFour{In the previous examples we showed the effectiveness of REDCHO against its non-robust version in \cite{aldana2021}. In the following we compare against other state of the art dynamic consensus methods, with particular focus on terminal precision for the EDC goal.} 
\begin{example}
\label{ex:comparison}
\RevAll{
In this example, we compare REDCHO with the Boundary-layer (B-layer) approach  from \cite{zhao2017} and the High-Order Linear protocol (HOL) from \cite{sen2020}. Both previous methods are able to achieve consensus  towards the average signal and its derivatives, but are not robust and require to share a whole vector between agents. In addition, we compare with the First-Order Linear (FOL) protocol in \cite{Solmaz2017} and the First-Order Sliding Mode (FOSM) protocol in \cite{freeman2019}. \RevFour{Both protocols are robust, but cannot obtain the derivatives of the average signal by construction. Thus, a robust exact differentiator \cite{levant2003} is applied locally at each agent to obtain derivatives of the average signal.}}

\RevTwo{In this setting, consider $\mathcal{G}$ constructed as a ring topology of $n=20$ agents.} \RevAll{Similarly as before, consider signals $u_i(t)=a_i\cos(\omega_it)$ where the $a_i,\omega_i$ are not shown for brevity. Note that all approaches can handle these type of signals, with at least a bounded terminal consensus error regardless of the order of the algorithm. An order of $m=2$ was used for REDCHO, B-layer and HOL, and an exact differentiator of order $m$ is applied for FOL and FOSM. Hence, all algorithms are able to obtain up to the second derivative of the average signal. All algorithms were implemented with parameters of similar magnitude, chosen such to roughly match same settling time for the sake of fairness. Moreover, we show the resulting consensus errors for all algorithms with $h = 10^{-6}$ as shown in Figure \ref{fig:h6} and $h=10^{-3}$ in Figure \ref{fig:h3} to show how they degrade as the discretization becomes coarser. \RevTwo{In addition, we simulated that agent 1 fails at $t=25$ and resets its state}, allowing us to evaluate robustness of the algorithms.}

\RevAll{As it can be observed, HOL and FOL methods have similar low precision in all cases before $t=25$. The reason is that neither of these methods are able to achieve exact convergence for sinusoidal signals. However, their performance does not degrade significantly when the time step is increased. On the other hand, it can be noted that the FOSM approach have better performance than the linear approaches when $h=10^{-6}$ due to its theoretically exact convergence. However, it degrades significantly when $h$ is increased as shown in Figure \ref{fig:h3}. The reason is that this method suffers from the chattering effect which is amplified for the higher order derivatives due to the exact differentiator. Note that the B-layer and REDCHO approaches have similar performance before $t=25$ with both sampling step sizes and outperform the other methods with at least one order of magnitude of precision improvement when $h=10^{-3}$ as shown in Figure \ref{fig:h3}. However, after $t=25$ both B-layer and HOL converge to consensus only up to a constant error due to their lack of robustness as shown by the $\|Y_0(t)-\bar{u}(t)\mathds{1}\|$ curves in both Figures \ref{fig:h6} and \ref{fig:h3}. Although other methods manage to recover from the failure of agent 1, REDCHO is the one with the best performance in all cases.}

\begin{figure}
    \centering
    \includegraphics[width=0.45\textwidth]{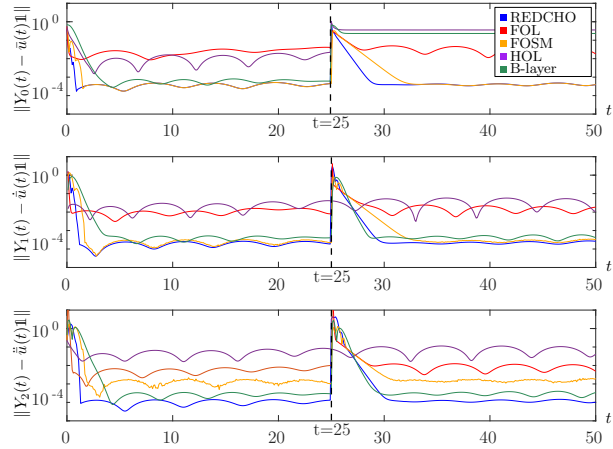}
\caption{Comparison of the magnitude of the average consensus errors for REDCHO, FOL, FOSM, HOL and B-layer in the case of a sampling step $h=10^{-6}$. }
    \label{fig:h6}
\end{figure}
\begin{figure}
    \centering
    \includegraphics[width=0.45\textwidth]{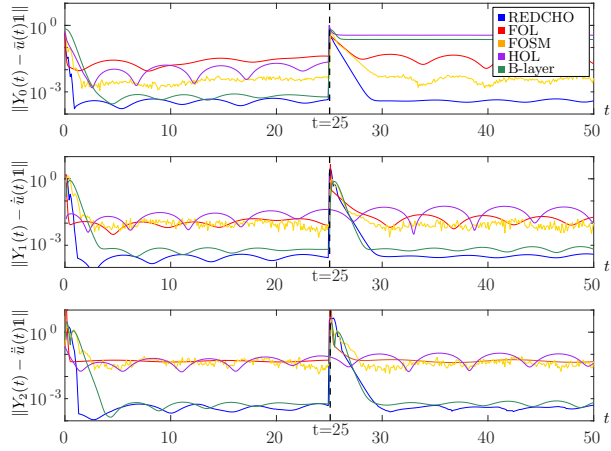}
\caption{Comparison of the magnitude of the average consensus errors for REDCHO, FOL, FOSM, HOL and B-layer in the case of a sampling step $h=10^{-3}$. }
    \label{fig:h3}
\end{figure}

\end{example}

\section{Conclusions}
In this work we proposed the REDCHO protocol. This new protocol achieves exact consensus towards the average of time varying signals and its derivatives distributed through a network. Proofs of convergence of the algorithm are given even when agents connect or disconnect from the network.  Simulation scenarios were designed to confirm the advantages of the proposed protocol. \RevTwo{Still, the proposed methodology works only when the changes in the network are isolated events. An analysis for general uncertain, time-varying  networks with persistent fast changes will be explored in future work.}

\appendix

{
\section{Convergence of consensus error of EDCHO}
\label{ap:edcho}
In this section, we provide an adaptation of \cite[Theorem 7]{aldana2021} focusing only in the dynamics of the consensus error $\tilde{Y}_\mu(t) = (\text{\normalfont I}_n-(1/n)\mathds{1}_n\mathds{1}_n^T)[y_{1,\mu}(t),\dots,y_{n,\mu}(t)]^T$ for the EDCHO protocol which are given by 
\begin{equation}
\begin{aligned}
\label{eq:main_algo_error2}
    \dot{\tilde{Y}}_{\mu}(t) &= \tilde{Y}_{\mu+1}(t) - k_\mu D\sgn{D^T\tilde{Y}_{0}(t)}{\frac{m-\mu}{m+1}}\\ &\text{for } 0\leq\mu \leq m-1, \\
    \dot{\tilde{Y}}_{m}(t) &=PU^{(m+1)}(t)- k_mD\sgn{D^T\tilde{Y}_0(t)}{0}
\end{aligned}
\end{equation}
where $P=(\text{\normalfont I}_n-(1/n)\mathds{1}_n\mathds{1}_n^T)$, $U(t) = [u_1(t),\dots, u_n(t)]^T$, $D$ is the incidence matrix of the communication graph $\mathcal{G}$ and some parameters $k_0,\dots,k_m>0$. Convergence to the origin for the consensus error is given in the following:
\begin{thm}(Adapted from \cite[Theorem 7]{aldana2021})
\label{prop:edcho}
Let $\mathcal{G}$ be a connected graph and $\tilde{Y}_\mu(t)\in\mathbb{R}^n$ under \eqref{eq:main_algo_error2}, subject to $\mathds{1}_n^T\tilde{Y}_\mu(t_0)=0$ for $0\leq \mu\leq m$. Moreover, let $L>0$ such that 
$
PU^{(m+1)}(t)\in[-L,L]^n 
$. Then, \RevFour{there exist gains} $k_0,\dots,k_m$ and $T>0$ such that $\tilde{Y}_\mu(t)=0, \forall \mu\in\{0,\dots,m\}, \forall t>t_0+T$.
\end{thm}
}
\vspace{-1cm}
\RevFour{\begin{rem}
\label{rem:levant}
Note that the gains $k_0,\dots,k_m$ from the previous result can be obtained using the parameters for the robust exact differentiator from \cite{levant2003} through a procedure described in \cite[Section 6]{aldana2021}.
\end{rem}
}
\section{Homogeneous differential inclusions}
\label{sec:homo}
In this section, we consider dynamical systems characterized by set-valued maps $F:\mathbb{R}^n\rightrightarrows\mathbb{R}^n$ instead of typical vector fields. Moreover, we assume some regularity conditions on such maps called the basic conditions. We say that a set valued map $F:G\rightrightarrows\mathbb{R}^n$ satisfies the basic conditions if for all $x\in G$ the set $F(x)$ is non-empty, bounded, closed, convex and the map $F$ is upper semi-continuous in $x$ \cite[Chapter 2.7]{filippov}.
The Filippov regularization for vector fields \cite{cortes2008}, commonly used to study discontinuous dynamical systems, satisfy the basic conditions by construction. In the following, let $\Delta_{\mathbf{r}}(\lambda) = \text{diag}([\lambda^{r_1},\dots,\lambda^{r_n}])$ where $\mathbf{r}=[r_1,\dots,r_n]$ are called the weights and $\lambda>0$. For any $x\in\mathbb{R}^n$, the vector
$
\Delta_{\mathbf{r}}(\lambda)x = [\lambda^{r_1}x_1,\dots,\lambda^{r_n}x_n]^T
$
is called its standard dilation (weighted by $\mathbf{r}$). The following are some definitions and results of interest regarding the so called $\mf{r}$-homogenety with respect to the standard dilation.

\begin{defn}[Homogeneous scalar functions]\cite[Definition 4.7]{bernuau2014}
A scalar function $V:\mathbb{R}^n\to\mathbb{R}$ is said to be $\mathbf{r}$-homogeneous of degree $d$  if
$
V(\Delta_{\mathbf{r}}(\lambda)x) = \lambda^dV(x)
$ for any $x\in\mathbb{R}^n$.
\end{defn}

\begin{defn}[Homogeneous set-valued fields]\cite[Definition 4.20]{bernuau2014}\label{def:homo}
A set-valued vector field $F:\mathbb{R}^n\rightrightarrows\mathbb{R}^n$ is said to be $\mathbf{r}$-homogeneous of degree $d$ if
$
F(\Delta_{\mathbf{r}}(\lambda)x) = \lambda^d\Delta_{\mathbf{r}}(\lambda)F(x)
$ for any $x\in\mathbb{R}^n$.
\end{defn}

\begin{prop}\cite[Lemma 4.2]{bernstein2005}\label{prop:homo_comparison}
Let $V_1,V_2:\mathbb{R}^n\to\mathbb{R}$ be continuous functions, $\mathbf{r}$-homogeneous of degrees $d_1>0,d_2>0$ (respectively). Moreover, let $V_1(x)$ to be positive definite. Then, $\forall x\in\mathbb{R}^n$,
$$
\beta_m V_1(x)^{d_2/d_1}\leq V_2(x)\leq \beta_M V_1(x)^{d_2/d_1}
$$
with $\beta_m=\inf\{V_2(x), \forall x: V_1(x)=1\}$ and $\beta_M=\sup\{V_2(x), \forall x: V_1(x)=1\}$.
\end{prop}
\begin{prop}\cite[Section 5]{bernstein2005}\label{prop:lie_homo}
Let $V:\mathbb{R}^n\to\mathbb{R}$ and $F:\mathbb{R}^n\rightrightarrows\mathbb{R}^n$ be a scalar field and set valued vector field, $\mathbf{r}$-homogeneous of degrees $l$ and $m$ respectively. Then, $L_FV(x)$ is $\mathbf{r}$-homogeneous of degree $l+m$.
\end{prop}
\begin{prop}\cite[Theorem 4.24]{bernuau2014}
\label{prop:converse}
Let $F:\mathbb{R}^n\rightrightarrows\mathbb{R}^n$ be $\mathbf{r}$-homogeneous of degree $m$ satisfying the standard assumptions. Moreover, assume that the differential inclusion $\dot{x}\in F(x)$ is strongly, globally asymptotically stable. Then, for any $k>\max(-m,0)$, there exists $V,W:\mathbb{R}^n\to\mathbb{R}$ continuously differentiable in all $\mathbb{R}^n$ and $\mathbb{R}^n\setminus\{0\}$ respectively. Moreover, $V$ is positive definite and $\mathbf{r}$-homogeneous of degree $k$ and $W$ is strictly positive outside the origin and $\mathbf{r}$-homogeneous of degree $k+m$. Finally, $\dot{V}\leq -W(x), \forall x\neq 0$.
\end{prop}
\bibliographystyle{plain}

\end{document}